\documentclass{pos}

\title{Search for Extra-dimensions in a single-jet and missing energy channel at CMS experiment}

\ShortTitle{Extra dimensions at CMS}

\author{\speaker{Leonardo Benucci}\\
        CMS Collaboration\\
        E-mail: \email{leonardo.benucci@cern.ch}}

\abstract{A possible solution to the hierarchy problem is the presence of extra spatial dimensions beyond the three ones which are known from our everyday experience. The phenomenological ADD model of large extra-dimensions predicts a missing transverse energy plus a single-jet signature. This contribution addresses the sensitivity of the CMS detector at the LHC pp collider to parameters of this model, focusing on the conditions expected for second half of 2010 running  ($\sqrt{s}=10$\,\TeV, O(100)\,\pbinv ). It is shown that a significant improvement of the existing limits can be obtained in such an early stage.}

\FullConference{XXth Hadron Collider Physics Symposium \\
		 November 16 -- 20, 2009 \\
		 Evian, France}

\usepackage{graphicx}

\newcommand{\GeV}{\,GeV\,}
\newcommand{\TeV}{\,TeV\,}

\newcommand{\pbinv}{pb$^{-1}$}

\newcommand{\PT}{p_T}

\newcommand{\stat}{{\rm (stat)}}
\newcommand{\syst}{{\rm (syst)}}
\newcommand{\MET}{\ensuremath{E_T^{miss}}\,}
\newcommand{\PYTHIA}{P{\footnotesize YTHIA}\,}
\newcommand{\SHERPA}{S{\footnotesize HERPA}\,}

\begin{document}

This contributions outlines the analysis procedures for the search of large extra dimensions in the missing transverse energy plus a single-jet channel, using the Compact Muon Solenoid (CMS) detector~\cite{bib:INTRO_CMS}. The description of the detector performance and the simulated samples of events correspond to what expected for 10\,\TeV center-of-mass energy and integrated luminosity up to 200\,\pbinv. Full details of the analysis can be found in Ref.~\cite{bib:our_PAS}.

The phenomenological ADD model~\cite{bib:INTRO_ADD1} aims to solve the hierarchy problem between the electroweak and Planck scales by introducing a number $\delta$ of extra spatial dimensions, which in the simplest scenario are
compactified over a torus and all have the same radius $R$. The fundamental scale $M_D$ is related to the effective $4$-dimensional Planck scale $M_{Pl}$  according to the formula $M_{Pl}^2\sim M_{D}^{\delta+2} R^{\delta}$. Current experimental constraints allow a scenario with $\delta \ge 2$, corresponding to extra-dimensions sizes below $5\cdot 10^{-2}$\,mm if the fundamental scale $M_D$ is of the order of TeV. Searches in both the jet+$\MET$ and the $\gamma+\MET$ have been performed by CDF~\cite{bib:INTRO_CDF}, D0~\cite{bib:INTRO_D0}, 
and LEP experiments~\cite{bib:INTRO_LEP}. The best 95\% confidence limits on $M_D$ are 1.40(1.04)\TeV for the extra dimensions scenario with $\delta = 2(4)$.

This study is focused on the production of a graviton $G$ balanced by a energetic hadronic jet via the $q\bar q \rightarrow gG$, $qg \rightarrow qG$, and $gg \rightarrow gG$ processes. The new physics signature is a high-transverse-momentum ($\PT>100\div 200$\,\GeV) jet in the central region of the detector, recoiling back-to-back in the transverse plane with a $\MET$ of similar magnitude.

The Standard Model process $Z(\nu \nu)+$jets leads to invisible energy recoiling against jets and is described by the same signature as the signal, thus the contribution from this ``irreducible'' background needs to be estimated in the best possible way. 
Other important background sources are $W +$jets with a leptonic $W$ decay (if the charged lepton is not reconstructed), QCD di-jets (when one or more jets are mismeasured and a significant amount of \MET is produced), and top-pair and single-top quark production, especially for events with few of collimated jets where leptons are not identified.

The ADD-model signal has been produced with the \SHERPA Monte Carlo generator~\cite{bib:GEN_Sherpa}, in different samples with $M_D$ ranging from 1 to 3\,\TeV and $\delta$ from 2 to 6. The transverse momentum of the outcoming parton was required harder than 150\,\GeV. A set of background processes were generated with a sample size corresponding to an integrated luminosity larger than 200\,\pbinv, then processed by a full simulation of the detector. The hadronization and fragmentation of quarks and gluons (along with the underlying event) were performed using \PYTHIA 6.409~\cite{bib:GEN_Pythia} and the CTEQ61L Parton Density Functions (PDF)~\cite{bib:SYST_CTEQ6M}
were used. With these production parameters, the signal cross sections at leading-order ranges from 279\,pb (for $M_D=$1\TeV, $\delta=2$) to 0.58\,pb (for $M_D=$3\TeV, $\delta=6$).

\begin{figure}[!Hbtp]
 \centering
  \includegraphics[width=0.40\textwidth]{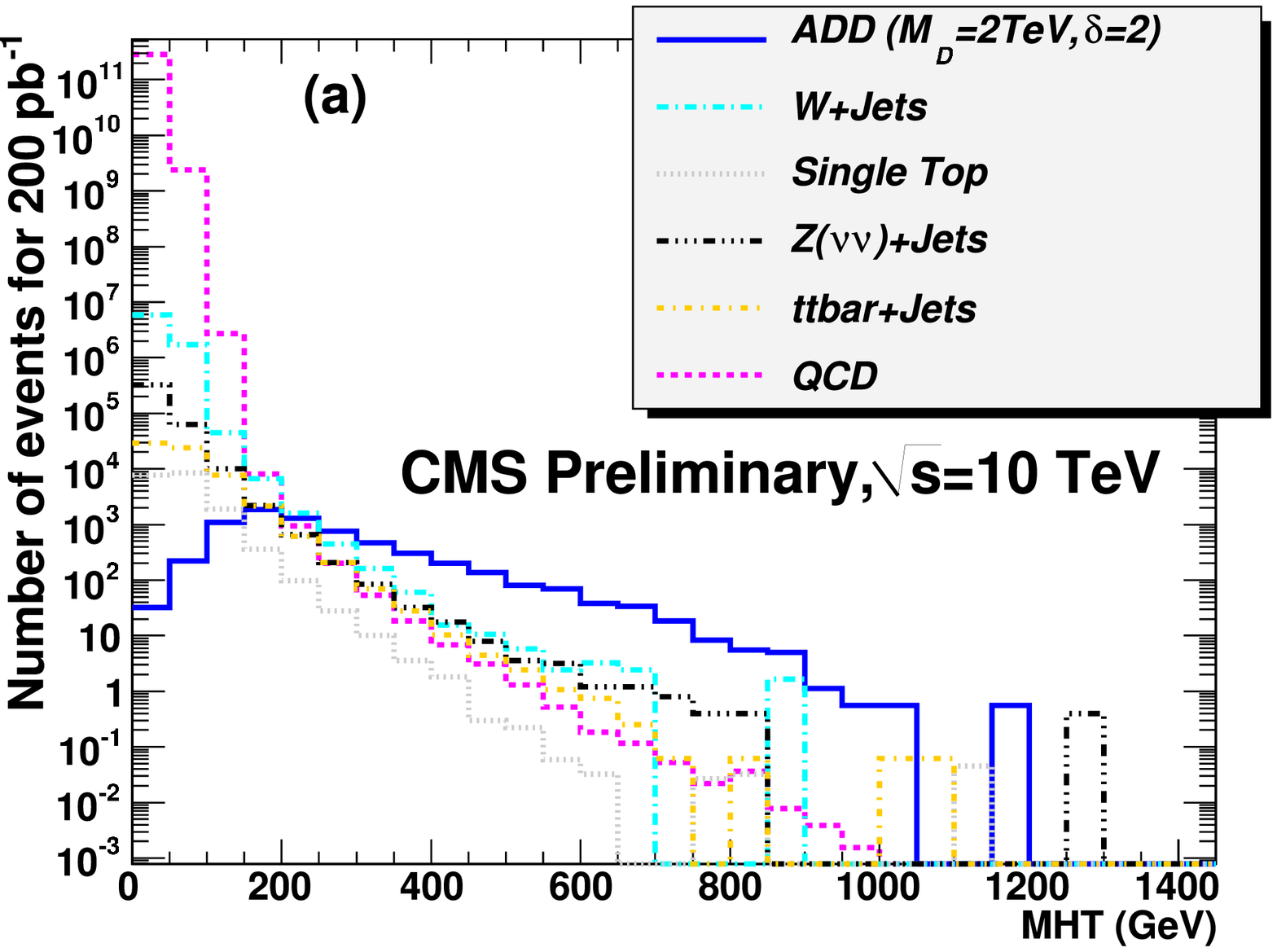}
  \includegraphics[width=0.40\textwidth]{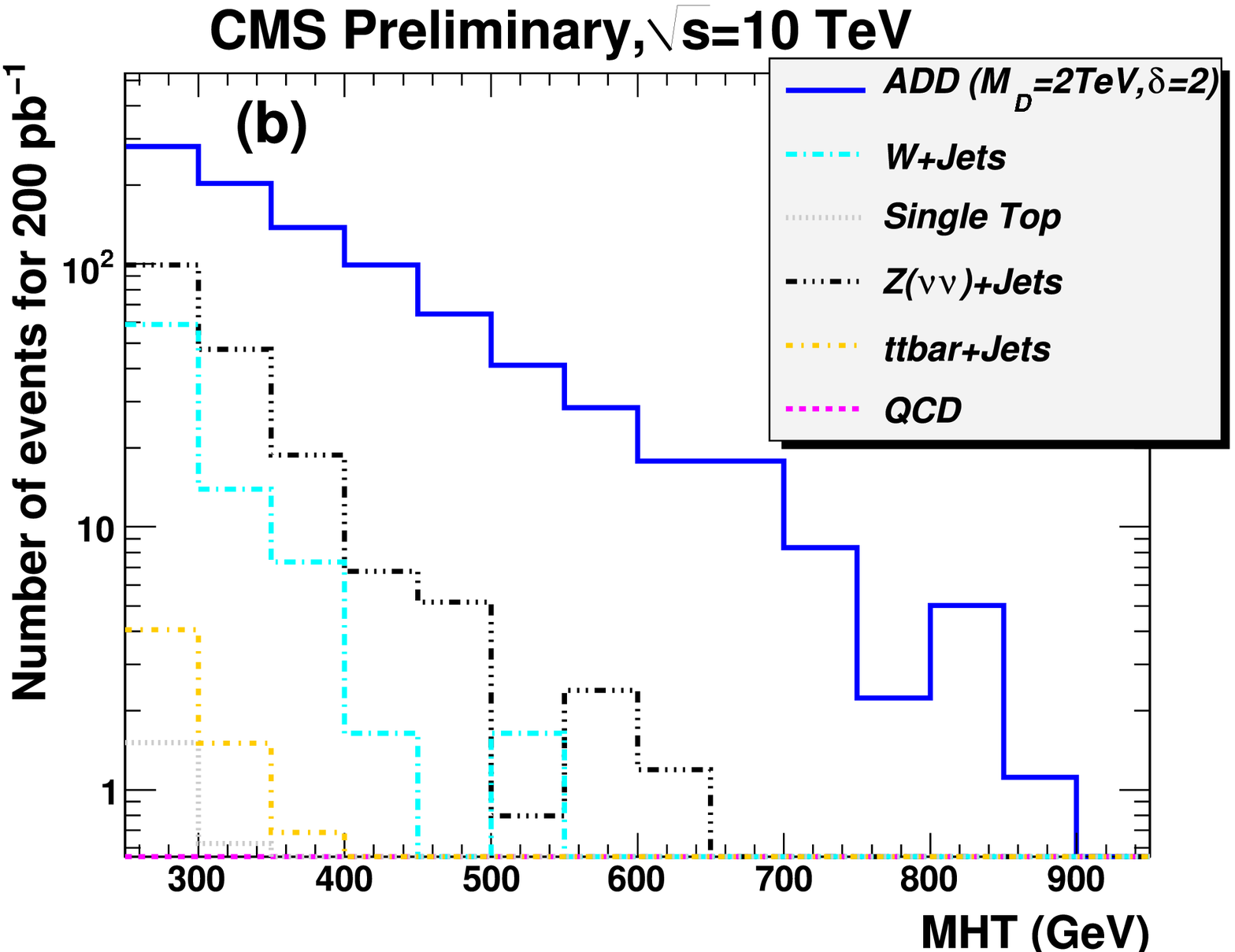}
  \caption{$(a)$ $MHT$ distributions for ADD signal ($M_D=2$\,\TeV, $\delta=2$) and relevant backgrounds before any selection, after 200\,\pbinv. $(b)$ Missing $H_T$ distribution after all selections are applied. Histograms are overlaid and number of events correspond to 200\,\pbinv. \label{fig:MHT_before_after}} 
\end{figure}

The analysis procedure is based on a set of cuts aimed to maximize the ratio of number of signal events over the square root of number of background events. At trigger level, a single jet stream is exploited, requiring at least 1 jet with $\PT>70(110)$\,\GeV at Level 1 (High Level Trigger). As shown in Fig.~\ref{fig:MHT_before_after}$(a)$, the signal leads to a long tail in the distribution of the vectorial sum of jets transverse momenta ($MHT$), hence a cut $MHT>$250 GeV was imposed at the pre-selection level. In order to reduce the impact of jets not coming from hard interaction, only jets with transverse momenta larger than 50\,\GeV within $|\eta|<3$ are considered. To clean the events from isolated lepton contamination, along with electrons and photons misidentified as jets, the fraction of jet energy collected by the electromagnetic calorimeter over the total energy is required to be lower than 0.9 and isolated tracks (having less than 10\% of $p_T$ in a ${0.02<\Delta R<0.35}$ cone) are removed. The leading jet is required to have $p_T>200$\GeV and $|\eta|<$1.7. A veto against events with more than two jets and a number of angular cuts $\Delta\phi(\textrm{jet\,1},MHT)>2.8$ and $\Delta\phi(\textrm{jet\,2},MHT)> 0.5$ complete the selection. The missing $H_T$ distribution for signal and background after all selections is shown in Fig.~\ref{fig:MHT_before_after}$(b)$. The signal shows up as an excess of events in addition to the dominant background $Z(\nu\nu)+$jets.

The most important systematic uncertainties from theory come from the cross section sensitivity to the renormalization and factorization scale ($^{+7.5\%}_{-6.7\%}$), and from uncertainties on the parton density function ($^{+11.5\%}_{-9.5\%}$). Uncertainties associated to energy and angular jet resolution (assumed to be 10\% and 0.1\,rad, respectively) and jet calibration (shifted by $\pm 10\%$) turn to be the most relevant instrumental effects. Their relative shift from the value with no systematic effect depends on the ADD points, ranging from 10\% to 16\%. The value of instantaneous luminosity, that is assumed to have a $\pm 10\%$ uncertainty, was incorporated.

The irreducible background of $Z(\nu\nu)+$jets (``invisible Z'') can be deduced from samples of events containing a high-$p_T$ $W$ boson decaying leptonically. The selection defining the control region has been kept the same of signal region,  except that a single isolated muon with $p_T > 20$\GeV and $|\eta|<$2.4 was required. This procedure allowed to reproduce the same kinematic region that was designed for the signal, but having a muon for which the hypothesis of coming from $W$ is highly probable. This sample was cleaned from the processes with at least one well-isolated muon passing the selection, then corrected by the ratio between $W+$jets and $Z+$jets production cross sections and muon reconstruction and isolation efficiency. The number of invisible $Z$ events in the signal region was found to be  $N(Z(\nu\nu)+{\rm jets})^{Sign}=163\pm 22\,\stat\pm 13\syst\pm 17{\rm (MC)}$\footnote{Notation $MC$ refers to the counting error due to the limited number of generated Monte Carlo events.}, to be compared with $N(Z(\nu\nu)+{\rm jets})^{MC}=182\pm 13\,\stat$.

The same region can be used to measure the $W(l\nu)+$jets contribution in the signal region. All the background estimates are consistent with the result from simulations. 
 	
\begin{figure}[!Ht]
 \centering
  \includegraphics[width=0.49\textwidth]{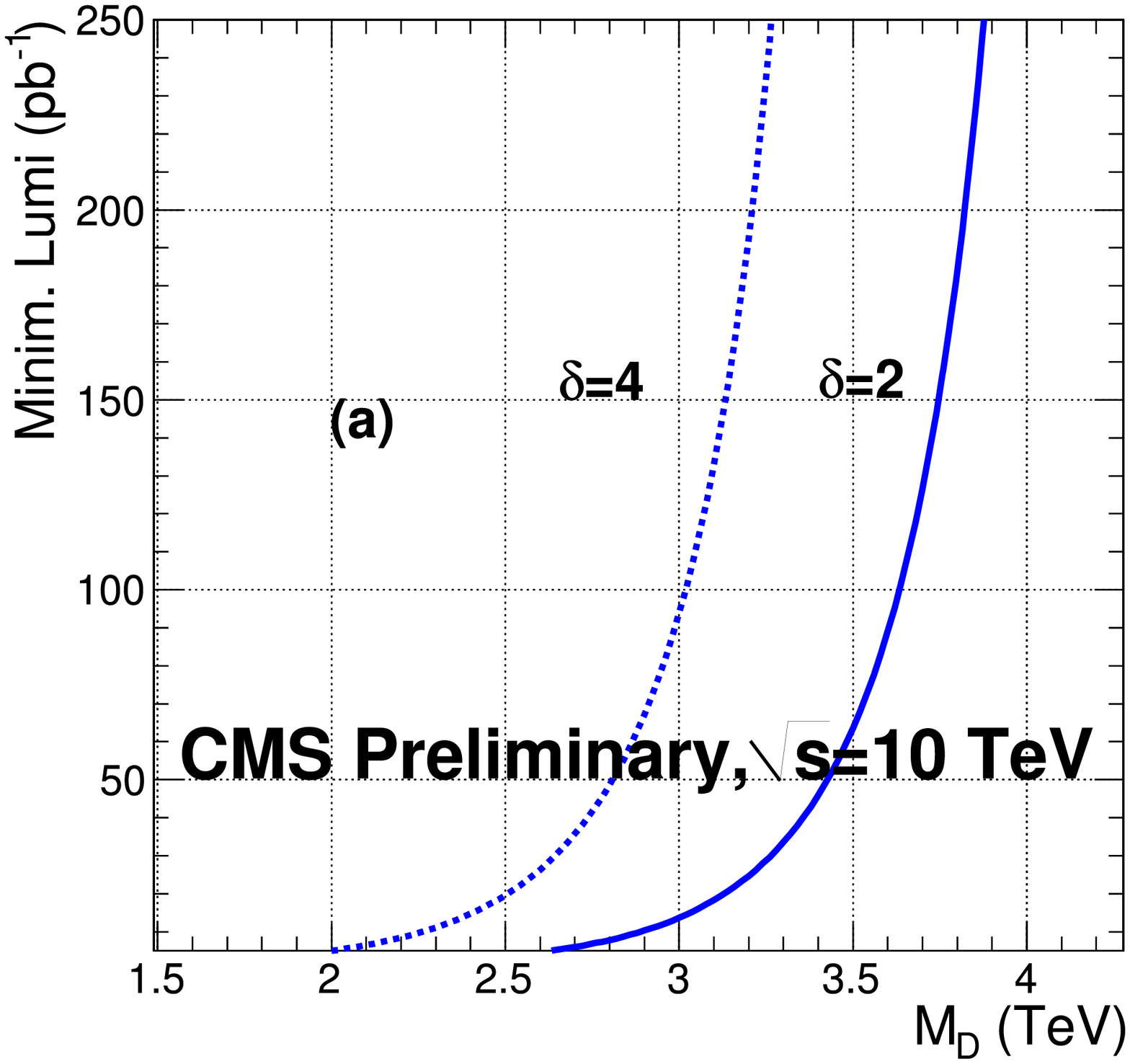}
  \includegraphics[width=0.49\textwidth]{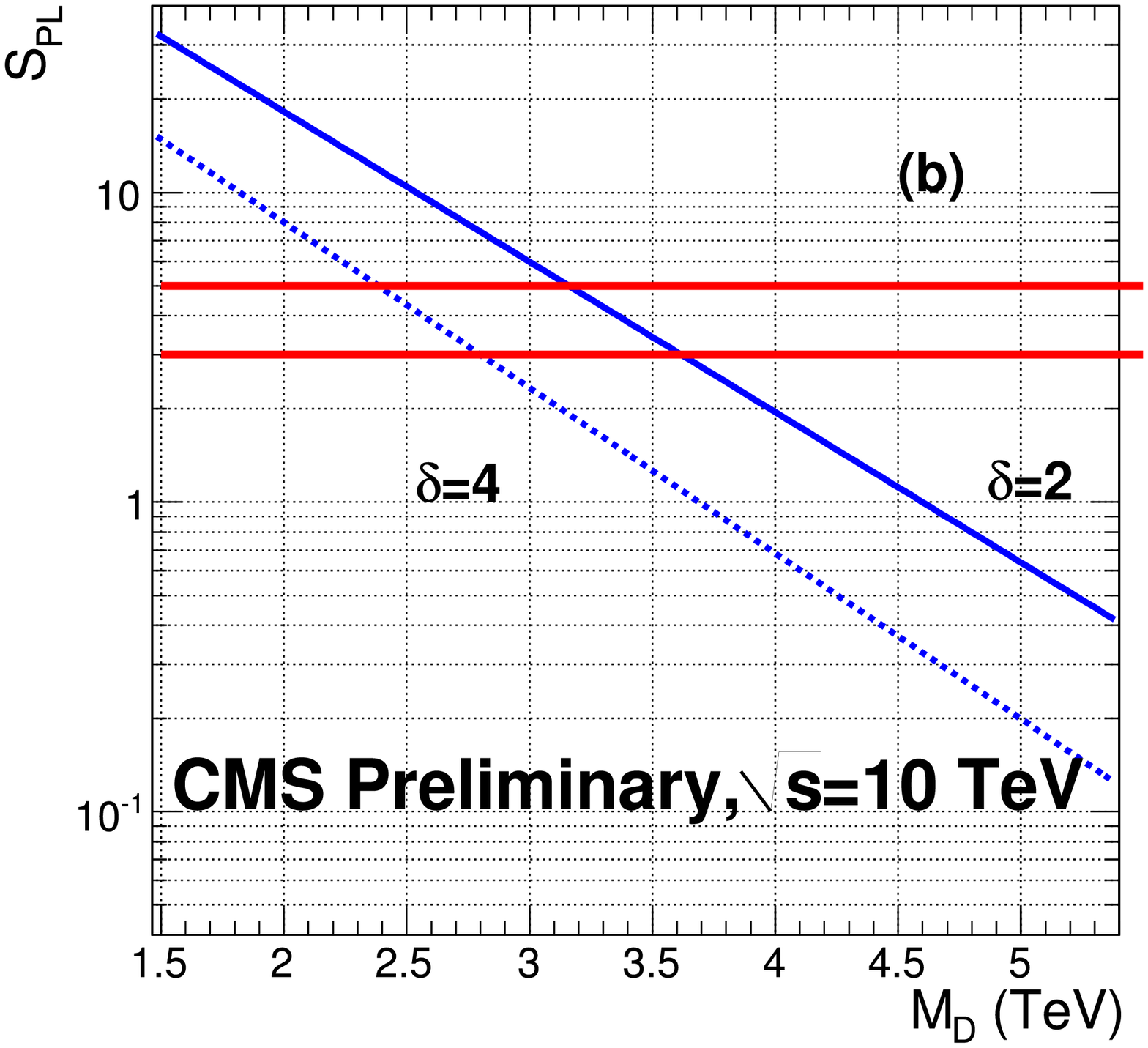}
  \caption{$(a)$: Exclusion plot at 95\% C.L., showing the minimum luminosity necessary to exclude a given value of $M_D$. $(b)$: Discovery potential of the analysis as a function of $M_D$ and $\delta$ after 200\,\pbinv. The horizontal thick lines correspond to 3$\sigma$ and 5$\sigma$ significance level. In both cases, sensitivity is plotted for two different extra dimension scenarios.} 
  \label{fig:STAT_Exclusion}
\end{figure}

The total background can be estimated as $N_B=243\pm 23\,\stat\pm 13\,\syst $ events after 200\,\pbinv of integrated luminosity. The 95\% C.L. limit was found by scanning the parameter space to minimize the negative Log Likelihood~\cite{bib:LIM_Cousins}. When different $M_D$, $\delta$ are interpolated, the exclusion plot in Fig.~\ref{fig:STAT_Exclusion}$(a)$ can be derived. The amount of data needed for a $5\sigma$ discovery was also calculated, using an estimator based on Profile Likelihood. When results from different signal points are interpolated, the plot in Fig.~\ref{fig:STAT_Exclusion}$(b)$ is obtained, that represents the sensitivity for a discovery after 200\,\pbinv.

These results indicate that the current exclusion limits from Tevatron experiments can be matched at LHC with the first physics run. Exclusion limits at 95\% for $M_D=3$\,\TeV, $\delta=2$, $M_D=2$\,\TeV, $\delta=4$ can be reached after only 11\,\pbinv and 5.0\,\pbinv, respectively; also early discoveries for $\delta=2(4)$ scenarios are possible, if $M_D$ is below 3.1(2.3)\,\TeV, respectively.

\end{document}